\documentclass[10pt, conference, letterpaper]{IEEEtran}
\IEEEoverridecommandlockouts
\usepackage{balance}
\usepackage{amssymb}
\usepackage{stmaryrd}
\usepackage{color}
\usepackage{bbm}
\usepackage{multirow}
\usepackage{graphicx,times}
\usepackage{epstopdf}
\usepackage{indentfirst}
\usepackage{CJK}
\usepackage{amsmath}
\usepackage{amsfonts}
\usepackage{txfonts}
\usepackage{mathrsfs}
\usepackage{subfigure}
\usepackage{graphicx}
\usepackage{theorem}
\usepackage{url}
\usepackage{microtype}
\usepackage{fancyhdr}
\usepackage{slashbox}

\begin{document}
	
\title{Forgery Attack Detection in Surveillance Video Streams Using Wi-Fi Channel State Information}
\author{\IEEEauthorblockN{Yong Huang, Xiang Li, Wei Wang,~\IEEEmembership{Senior Member,~IEEE}, Tao Jiang,~\IEEEmembership{Fellow,~IEEE}, Qian Zhang,~\IEEEmembership{Fellow,~IEEE}}
	\thanks{Part of this work has been presented at IEEE INFOCOM 2021~\cite{yong2021towards}.}
	\thanks{This work was supported in part by the National Key R\&D Program of China under Grant 2019YFB1803400, National Science Foundation of China with Grant 62071194, RGC under Contract CERG 16204418, Contract 16203719, 16204820, R8015, and in part by the Guangdong National Science Foundation under Grant 2017A030312008. \textit{(Corresponding author: Wei Wang.)}}
	\thanks{Y. Huang, W. Wang and T. Jiang are with the School of Electronic Information and Communications, Huazhong University of Science and Technology, Wuhan 430074, China (e-mail:\{yonghuang, weiwangw, taojiang\}@hust.edu.cn).}
	\thanks{X. Li is with Department of Electrical and Computer Engineering, Carnegie Mellon University, Pittsburgh 15213, United States (e-mail:xl6@andrew.cmu.edu).}
	\thanks{Q. Zhang is with Department of Computer Science and Engineering, Hong Kong University of Science and Technology, Clear Water Bay, Hong Kong, China (e-mail:qianzh@cse.ust.hk).}}

\maketitle

\begin{abstract}
The cybersecurity breaches expose surveillance video streams to forgery attacks, under which authentic streams are falsified to hide unauthorized activities. Traditional video forensics approaches can localize forgery traces using spatial-temporal analysis on relatively long video clips, while falling short in real-time forgery detection. The recent work correlates time-series camera and wireless signals to detect looped videos but cannot realize fine-grained forgery localization. To overcome these limitations, we propose Secure-Pose, which exploits the pervasive coexistence of surveillance and Wi-Fi infrastructures to defend against video forgery attacks in a real-time and fine-grained manner. We observe that coexisting camera and Wi-Fi signals convey common human semantic information and forgery attacks on video streams will decouple such information correspondence. Particularly, retrievable human pose features are first extracted from concurrent video and Wi-Fi channel state information (CSI) streams. Then, a lightweight detection network is developed to accurately discover forgery attacks and an efficient localization algorithm is devised to seamlessly track forgery traces in video streams. We implement Secure-Pose using one Logitech camera and two Intel 5300 NICs and evaluate it in different environments. Secure-Pose achieves a high detection accuracy of 98.7\% and localizes abnormal objects under playback and tampering attacks.
\end{abstract}

\begin{IEEEkeywords}
Surveillance system, forgery detection and localization, cross-modal learning
\end{IEEEkeywords}

\section{Introduction}
\begin{figure}
	\centering
	\includegraphics[width=0.87\linewidth]{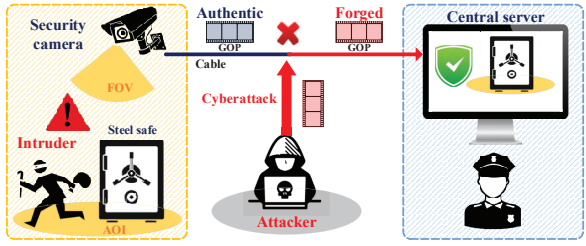}
	\caption{Illustration of video forgery attacks in surveillance systems.}
	\label{fig:attackscenario}
\end{figure}

With the increasing needs of safety and security in our daily life, video surveillance systems have gained a lot of traction in many indoor applications, such as crime prevention in banks and customer monitoring in retail stores~\cite{liu2013intelligent,zhong2017multi}. As their popularity and prominence rapidly grow in the physical world, these systems inevitably become attractive attack surfaces in the cybersecurity space. Recent studies have demonstrated that attackers can infiltrate into the surveillance system by exploiting vulnerabilities of the monitoring camera~\cite{exploiting2013} or hijacking its connection Ethernet cables~\cite{looping2015} and then alter the authentic live video streams to conceal illegal human activities in the monitored area without showing any perceptible clues in the central server's screen as shown in Fig.~\ref{fig:attackscenario}. 

Although extensive efforts have been devoted to defending against video forgery attacks, existing approaches still fall short in achieving both real-time forgery detection and fine-grained forgery localization. Traditional watermark-based approaches require dedicated modules on security cameras for video integrity preservation, while not all camera manufacturers support such modules~\cite{fayyaz2020improved}. Alternatively, many video forensics approaches that exploit video statistic characteristics~\cite{yang2016using,chen2015automatic,ulutas2017frame,wang2007exposing} are developed to detect tampered frames and further localize forgery traces. However, these approaches rely on spatial-temporal analysis on relatively long video clips, which are ill-suited for live video streams in time-critical surveillance systems. Additionally, the recent work~\cite{lakshmanan2019surfi} demonstrates that Wi-Fi signals can be leveraged to expose video looping attacks on surveillance systems. However, it employs handcrafted event-level timing and frequency features from time-series Wi-Fi and camera signals, which leads to a long response time and cannot realize fine-grained forgery localization. Hence, none of existing approaches simultaneously satisfies real-time and fine-grained requirements of forgery detection and localization in video surveillance systems.

The pervasive coexistence of surveillance cameras and Wi-Fi devices offers the opportunity to thwart video forgery attacks in a real-time and fine-grained manner. Nowadays, many areas under surveillance cameras, such as shops and homes, are also covered by Wi-Fi hotspots to provide us ubiquitous wireless connectivity~\cite{wang2019cross,wang2020enabling}. In such areas, not only visible light but also Wi-Fi signals interact with involving human objects, because human bodies act as reflectors in the Wi-Fi frequency range. In this condition, camera and Wi-Fi signals convey the common human semantic information. If forgery attacks are launched on camera signals, such cross-modal information correspondence will be decoupled, which can be exploited for timely forgery detection and accurate forgery localization.   

Toward this end, we present Secure-Pose, a novel cross-modal system that effectively detects and localizes forgery traces in live surveillance videos using coexisting Wi-Fi signals. Our key insight is that channel state information (CSI) measurements also convey human semantic features and video forgery attacks will mismatch them with those of video streams. Thus, Secure-Pose can exploit CSI measurements for forgery detection and localization on suspicious live video streams.

To realize the above idea, we have to address the following three challenges.

\textit{1) How to synchronize noisy CSI measurements with video frames?} Generally, raw CSI measurements have variable time intervals and contain massive noisy components, which significantly hampers periodicity of CSI measurements as well as their information correspondence with video streams. To deal with this issue, we first use linear interpolation to resample a group of fixed-interval CSI samples for each video frame. Then, we remove high-frequency noise in CSI measurements with a low-pass Butterworth filter. 

\textit{2) How to effectively disentangle human semantic features from CSI measurements?} Due to a low spatial resolution, recovering complete video frames from CSI measurements is a highly ill-posed problem. To relieve this problem, we design a cross-modal learning network, namely CSI2Pose, that exploits spatial and temporal information in CSIs to estimate limb-level human pose features, i.e., Joint Heat Maps (JHMs) and Part Affinity Fields (PAFs). Additionally, two element-wise weighted training losses are devised to train CSI2Pose network with the guidance of OpenPose network.

\textit{3) How to efficiently detect and localize forgery traces in live video streams?} Since multiple people would appear under the camera, it is computationally inefficient to perform person-by-person forgery detection and localization in video frames. To avoid this issue, we develop a lightweight detection network that compresses sparse JHM tensors into single-channel feature planes for fast forgery detection. Once detected, the estimated JHMs and PAFs are reasonably fused to generate suspicious body keypoints and localize abnormal objects in forged regions.

\textbf{Summary of Results.} We implement Secure-Pose using a Logitech 720p camera and two Intel 5300 NICs, and evaluate it in different environments. Secure-Pose achieves a high detection accuracy of 98.7\%. Moreover, it can successfully recognize 99.2\% of forged video streams and mistakenly classify just 1.8\% authentic ones. In addition, Secure-Pose can localize body keypoints of forged human objects with a mPCK@0.25 of 78.4\% under various forgery attacks.

\textbf{Contributions.} First, we propose Secure-Pose that exploits coexisting camera and Wi-Fi signals to reliably detect and localize forgery attacks in live video streams. Second, we design an effective cross-modal network to project Wi-Fi signals into visual semantic features and devise efficient approaches for fast attack detection and forgery localization. Third, we implement Secure-Pose using one Logitech 720p camera and two Intel 5300 NICs and evaluate it in different environments to verify its effectiveness against various forgery attacks.

\section{Related Work}\label{sec:related work}
\textbf{Video Surveillance Systems.} Nowadays, many surveillance systems can automate surveillance tasks that were once performed by humans~\cite{liu2013intelligent}, and they perform an automatic understanding of on-going scenes via advanced computer vision techniques, such as object detection, recognition, tracking~\cite{cao2019openpose,he2017mask}. However, the emerging cyberattacks demonstrate that surveillance systems can be easily forged by exploiting the vulnerability of a surveillance camera~\cite{exploiting2013} or hijacking a connecting Ethernet cable between the camera and the central server~\cite{looping2015}, which renders video contents unreliable. 

\textbf{Video Forgery Detection.} Passive video forensics approaches leverage video statistic characteristics to discover forgery traces~\cite{fayyaz2020improved,yang2016using,chen2015automatic,ulutas2017frame,wang2007exposing}. However, such approaches generally rely on relatively long video clips and are ill-suited for live video feeds. The recent work~\cite{lakshmanan2019surfi} compares event-level timing and frequency information from Wi-Fi and camera signals to detect camera looping attacks. However, this work cannot realize timely forgery detection and fine-grained forgery localization. 

\textbf{Human Perception Using RF Signals.} Recent years have witnessed much progress in human-oriented RF sensing. The work~\cite{yong2020authenticating} utilizes on-body RF signals to detect wearable devices. The authors in~\cite{zhao2018rf} use a dedicated FMCW radio to capture 3D human skeletons from RF signals. Relying on commercial Wi-Fi devices, WiPose~\cite{jiang2020towards} is proposed to reconstruct 3D human skeletons in the single-person scenario. Moreover, Person-in-WiFi~\cite{wang2019person} is designed to learn body segmentation mask and joint coordinates from Wi-Fi signals. 

\section{Threat Model and Wi-Fi CSI Signatures}\label{sec:motivation}

\subsection{Forgery Attacks on Live Surveillance Videos}
We focus on a surveillance system, which consists of fixed security cameras and one remote central server as shown in Fig.~\ref{fig:attackscenario}. For monitoring an area of interest (AOI) in real time, one security camera is responsible for capturing a group of pictures (GOP) that generally lasts for several seconds and immediately transmitting them to the central server~\cite{wang2015resource}. Under the camera’s field of view (FOV), human objects are the main targets for behavior monitoring and trace tracking in many indoor environments~\cite{liu2013intelligent}, such as banks and retail stores. Moreover, we make no assumptions on the number of people as well as their motions under the camera’s FOV. 

To conceal illegal activities, an attacker can first penetrate into the surveillance system via launching cyberattacks, such as hijacking security cameras~\cite{exploiting2013} or Ethernet cables~\cite{looping2015}, and then alter genuine video streams. Since live video streams consist of successive GOPs that are separately transmitted, one GOP is the basic unit for launching forgery attacks. The penetrated attacker can trigger playback and tampering attacks on each intercepted GOP. In playback attacks, the attacker replaces authentic GOPs with previously-recorded ones. In tampering attacks, the attacker removes or inserts some human objects in authentic GOPs. The received GOPs by the central server that contain mismatching activities with respect to authentic ones are considered as forged. In forged frames, the erased or added people are referred to as abnormal objects.  

Furthermore, we assume that a pair of Wi-Fi transceivers are co-located with the security camera. Since it is extremely difficult to mimic complicated Wi-Fi signals based on video contents, CSI measurements from the neighboring Wi-Fi devices are considered to be authentic. 

\begin{figure}
	\centering
	\includegraphics[width=0.95\linewidth]{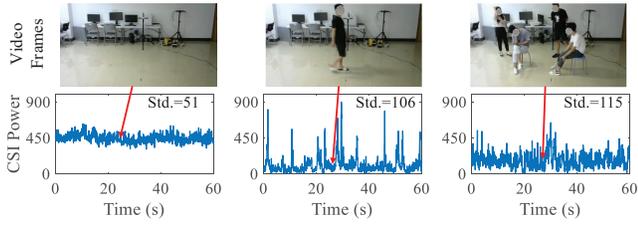}
	\caption{Concurrent CSI measurements and video frames.}
	\label{fig:feasibilitystudy}
\end{figure}

\subsection{CSI Signatures in Coexisting Wi-Fi Signals}
Channel state information describes how a Wi-Fi signal propagates between a transceiver pair for characterizing current channel conditions. Since the human body is a good reflector at the Wi-Fi frequency range~\cite{adib2015capturing}, Wi-Fi signals will have rich interactions with body limbs. According to the CSI-motion model~\cite{wang2015understanding}, the CSI at time $ t $, frequency $ f $ and antenna $ i $ can be expressed as a sum of static and dynamic channel frequency responses (CFRs), which is given by
\begin{align}\label{eq:csi model}
\mathbf{H}(t,f,i) = \left( \mathbf{H}_s(f,i)+ \mathbf{H}_d(t,f,i) \right) e^{-j\rho(t,f,i)}. 
\end{align}
Therein, the static CFR $ \mathbf{H}_s(f,i) $ is the sum of responses for static paths. The dynamic CFR $ \mathbf{H}_d(t,f,i) $ represents the sum of responses for paths with time-varying lengths caused by human movements. Moreover, the phase shift $ \rho(t,f,i) $ is unknown offsets caused by hardware imperfections. As the above equation indicates, resembling to visible light, Wi-Fi CSI measurements can be also affected by human bodies. Hence, information correspondence about human objects exists among coexisting video and Wi-Fi signals in AOI.

To verify this, we collect concurrent CSI and video streams in one office room. As depicted in Fig.~\ref{fig:feasibilitystudy}, when no people is within the targeted area, CSI measurements have a high power level and are relatively static. However, once people are present, received CSI power fluctuates dramatically as people change their positions and poses. The above observations are due to that the human body is a good reflector at the Wi-Fi frequency range, and therefore their presence near the Wi-Fi transceiver would incur fast changes of the dynamic CFR $ \mathbf{H}_d $ in Eq.~\eqref{eq:csi model}. Note that Fig.~\ref{fig:feasibilitystudy} demonstrates only one subcarrier of CSI measurements. According to the IEEE 802.11n/ac/ax Wi-Fi protocols, both OFDM and MIMO technologies are adopted for high-throughput transmission. Thus, at time $ t $, one CSI measurement could have hundreds of entries that probe human objects over different antennas and frequencies. Hence, similar to visible light captured by a camera, Wi-Fi CSI measurements can also convey rich human body information, which can be leveraged to check the veracity of video streams under forgery attacks.

\section{System Design}\label{sec:design}

\begin{figure}
	\centering
	\includegraphics[width=0.97\linewidth]{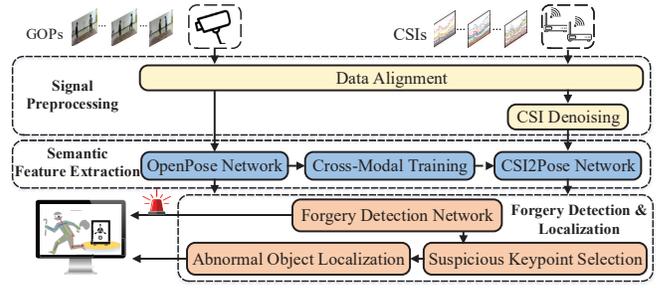}
	\caption{System architecture of Secure-Pose. }
	\label{fig:systemflow}
\end{figure}

\subsection{System Overview}

As depicted in Fig.~\ref{fig:systemflow}, Secure-Pose takes concurrent video streams and Wi-Fi signals as input to detect and localize forgery traces in live surveillance videos. Secure-Pose can be a part of the surveillance system, and its properties empower the system to quickly alarm on-going cyberattacks and seamlessly track potential intruders. The core of Secure-Pose includes the following three components. 

\begin{itemize}
	\item  \textbf{Signal Preprocessing.} Taking camera and Wi-Fi signals as input, our system first aligns CSI power measurements with video frames via linear interpolation for better data accordance. Then, it removes high-frequency noise in CSI signals via a low-pass Butterworth filter.
	\item  \textbf{Semantic Feature Extraction.} In this component, our system utilizes a well-designed CSI2Pose network that takes advantages of 3D convolutions to extract sparse JHMs and PAFs from CSI sequences. Then, a cross-modal training scheme is devised to effectively train CSI2Pose network with the supervision of OpenPose network.
	
	\item  \textbf{Forgery Detection and Localization.} Based on visual and wireless JHMs and PAFs, our system exploits a lightweight network to detect forgery attacks in each received GOP. Once detected, our system efficiently fuses the semantic features and fast localizes abnormal objects in each forged frame. 
\end{itemize}

\subsection{Signal Preprocessing}\label{subsection: signal preprocessing}


For video streams, we denote the decoded video frames in one received GOP as $ \mathcal{I} = \left\lbrace \mathbf{I}^{1}, \cdots, \mathbf{I}^{m}, \cdots,  \mathbf{I}^{ M}\right\rbrace  $, where $ M $ represents the GOP size and $ \mathbf{I}^{m} \in \mathbb{R}^{H\times W \times 3} $ is a complete RGB image. Therein, $ H $ and $ W $ indicate the frame height and width, respectively. For Wi-Fi signals, we take time-series CSI power measurements as input to remove the impact of unknown phase offsets as aforementioned. Concretely, let $ N_{t} $ and $ N_{r} $ be the numbers of transmitting and receiving antennas, respectively, $ K $ the number of OFDM subcarriers. The received CSI power sequence corresponding to $ \mathcal{I} $ can be expressed as $ \mathcal{A} = \left\lbrace  \mathbf{A}^{1}, \cdots, \mathbf{A}^{n}, \cdots, \mathbf{A}^{N}  \right\rbrace $, where $ \mathbf{A}^{n} \in \mathbb{R}^{N_{t} N_{r} \times K}  $ is one CSI power measurement and $ N $ is the measurement number. 

\begin{figure}
	\centering
	\subfigure[Count of CSIs per video frame.]{
		\includegraphics[width=0.475\linewidth]{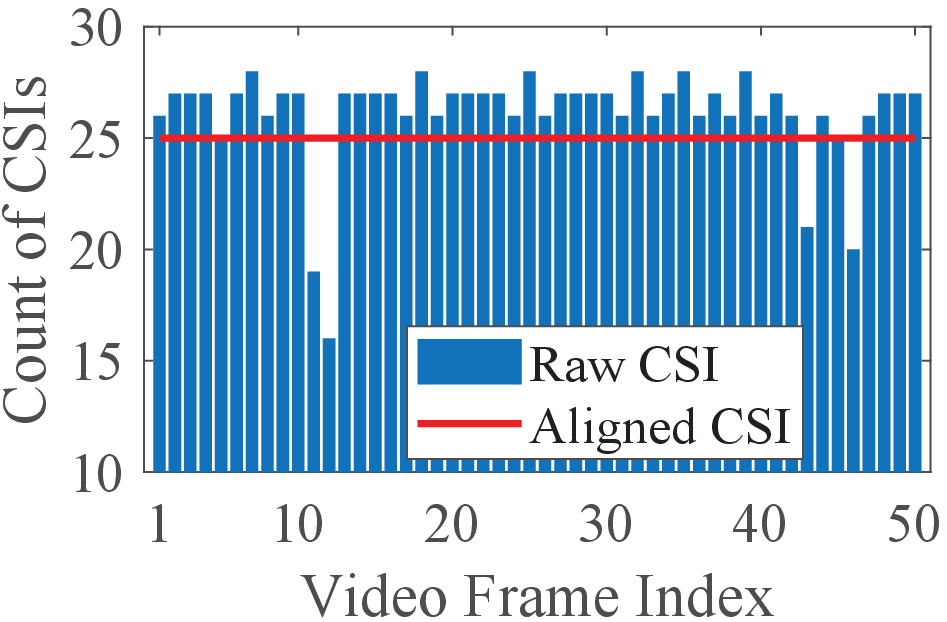}}
	\label{2a}
	\subfigure[CSIs w/o Butterworth filtering.]{
		\includegraphics[width=0.475\linewidth]{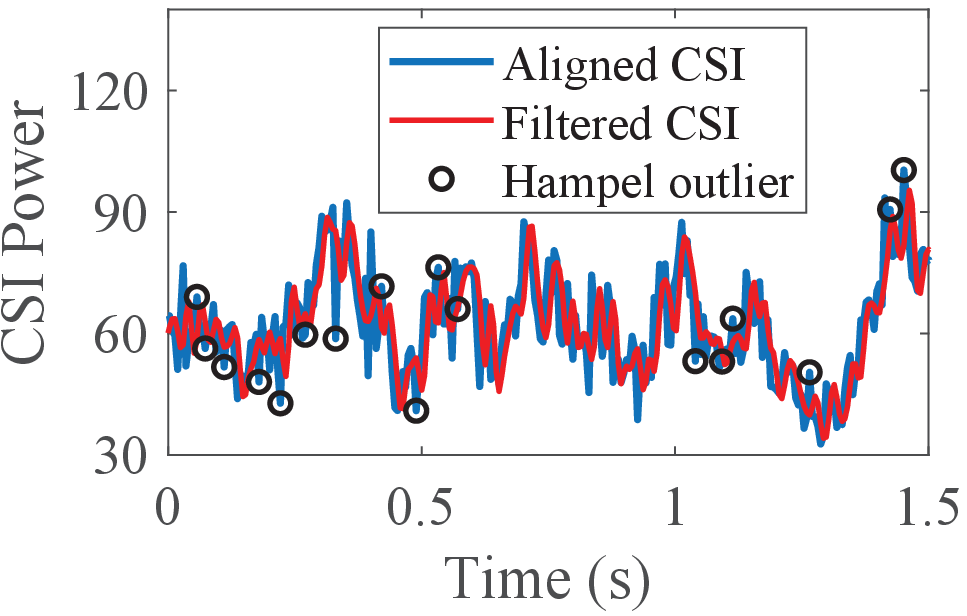}}
	\label{2b}\\
	\caption{CSI measurements alignment and Butterworth filtering.}
	\label{fig:datapreprocessing} 
\end{figure}

\textbf{Data Alignment.} Since our system relies on concurrent camera and Wi-Fi signals for attack detection, the synchronization between two data modalities is critical. Let us assume that the camera has a FPS, i.e., frames per second, of $ F_{I} $ Hz and the Wi-Fi receiver has a CSI sampling rate of $ F_{W} $ Hz. Because $ F_{W} $ is basically much larger than $ F_{I} $, we can sequentially correspond $ F \le   F_{W}/F_{I}  $ CSI measurements to each video frame. However, due to the random access protocol as well as packet loss, the number of CSI samples that fall between any two successive video frames is variable as shown in Fig.~\ref{fig:datapreprocessing}(a), which consequently weakens their periodicity. 

To deal with this issue, Secure-Pose aligns CSI measurements $ \mathcal{A} $ with video frames $ \mathcal{I} $ by resampling $ F $ CSI measurements for each video frame $ \mathbf{I}^{m} $ based on their timestamps. Mathematically, we denote $ t_{m-1} $ and $ t_m $ as the timestamps of successive video frames $ \mathbf{I}^{m-1}$ and $ \mathbf{I}^{m} $, respectively. For the video frame $ \mathbf{I}^{m} $, our system resamples a set of $ F $ CSI measurements $ \left\lbrace \mathbf{\tilde{A}}^{1}, \cdots, \mathbf{\tilde{A}}^{f}, \cdots, \mathbf{\tilde{A}}^{F}  \right\rbrace  $ with a constant time interval $ \Delta t = (t_{m}-t_{m-1})/F $ using low-complexity linear interpolation, and thus each resampled measurement $ \mathbf{\tilde{A}}^{f} $ can be obtained by $  $
\begin{align} 
\mathbf{\tilde{A}}^{f} = \mathbf{A}^{n-1} + \eta \cdot (\mathbf{A}^{n}-\mathbf{A}^{n-1}),
\end{align}
where $ t_{n-1} \leq t_f = t_{m-1} + f \Delta t \leq t_{n} $ and $ 0 \leq \eta = \frac{t_{f} - t_{n-1}}{t_{n} - t_{n-1}} \leq 1 $. As depicted in Fig.~\ref{fig:datapreprocessing}(a), the aligned CSI measurements show a constant periodicity and thus have better information accordance with video frames.

\textbf{CSI Denoising.} After data alignment, we denote the CSI power sequence corresponding to $ t $-th transmitting antenna, $ r $-th receiving antenna and $ k $-th subcarrier as $ \tilde{\mathbf{a}}_{t,r,k} = \left\lbrace  \tilde{a}_{t,r,k}^{1} \cdots , \tilde{a}_{t,r,k}^{n},\cdots,\tilde{a}_{t,r,k}^{MF} \right\rbrace $, where $  \tilde{a}_{t,r,k}^{n} \in \mathbb{R} $ is a CSI power measurement. As shown in Fig.~\ref{fig:datapreprocessing}(b), the aligned CSI measurements still consist of many impulse and burst noise incurred by transceiver hardware imperfections and environmental interference~\cite{ma2019wifi}. Since the fluctuations caused by human objects are relatively slow~\cite{wang2016gait}, it is necessary to denoise uncorrelated high-frequency noise. Thus, for $ \tilde{\mathbf{a}}_{t,r,k} $, we adopt a low-pass Butterworth filter with a cutoff frequency at 60 Hz for its maximally flat response within the pass-band. As Fig.~\ref{fig:datapreprocessing}(b) depicts, compared with Hampel identifier used in our previous work~\cite{yong2021towards}, the denoised CSI measurements using the Butterworth filter are smoother and contain less fluctuations that are irrelevant to human objects. 

Towards this end, for each video stream $ \mathcal{I} $, Secure-Pose outputs a group of resampled and refined CSI power measurements as $ \mathcal{R} = \left\lbrace  \mathbf{R}^{1}, \cdots, \mathbf{R}^{m}, \cdots, \mathbf{R}^{M} \right\rbrace  $, where $ \mathbf{R}^{m} \in \mathbb{R}^{N_{t} N_{r} \times K \times F} $ is the RF frame that is synchronized to the video frame $ \mathbf{I}^{m} $.

\subsection{Semantic Feature Extraction} 


\textbf{Semantic Feature Representation.} Generally, bounding boxes, segmentation masks and body keypoints could be candidate semantic features for human perception. Specifically, bounding boxes are widely used in object recognition and can indicate the locations of persons under the camera's FOV~\cite{ren2017faster}. Moreover, segmentation masks are sets of binary pixels to sketch the contours of human instances in the task of instance segmentation~\cite{he2017mask}. Though the above two features may be leveraged for detecting abnormal objects, they cannot provide pose information for tracking intruders' actions. Differing from them, body keypoints characterize human poses using 2D sparse anatomical points in the task of pose estimation~\cite{cao2019openpose}. Such limb-level features convey the locations as well as motion states of human objects and are also perceptible by Wi-Fi signals as human limbs act as reflectors at the Wi-Fi frequency range~\cite{adib2015capturing}. Based on the above reasons, we formulate the semantic feature extraction as a human pose estimation problem.

We adopt Joint Heat Maps and Part Affinity Fields to represent body keypoints for their effectiveness in human pose estimation~\cite{cao2019openpose}, and extract them from concurrent video and RF frames, respectively. To do this, our system uses the Body-14 model that describes the human body as $ J=14 $ skeleton keypoints, i.e., nose, neck, shoulders, elbows, wrists, hips, knees and ankles, which accordingly yield $ C=13 $ body limbs. Concretely, given a video frame $\mathbf{I}^{m} $, JHMs indicate the confidence maps of keypoint locations in the image space and can be represented by a 3D tensor as $ \mathbf{S}^{m} = \left\lbrace  \mathbf{s}_1^m, \mathbf{s}_2^m, \cdots, \mathbf{s}_J^m \right\rbrace \in \mathbb{R}^{H\times W \times J} $, where $ \mathbf{s}_j^m \in \mathbb{R}^{H\times W} $ is the 2D confidence map of $ j $-th keypoint. Moreover, PAFs encode location and orientation information of body limbs and can be denoted as a 4D tensor as $ \mathbf{L}^m = \left\lbrace  \mathbf{l}_1^m, \mathbf{l}_2^m, \cdots, \mathbf{l}_C^m  \right\rbrace \in \mathbb{R}^{H\times W \times 2 \times C} $, where $ \mathbf{l}_c^m  \in \mathbb{R}^{ H\times W \times 2 }$ is a 2D vector field for $ c $-th limb.

\begin{figure}
	\centering
	\includegraphics[width=0.99\linewidth]{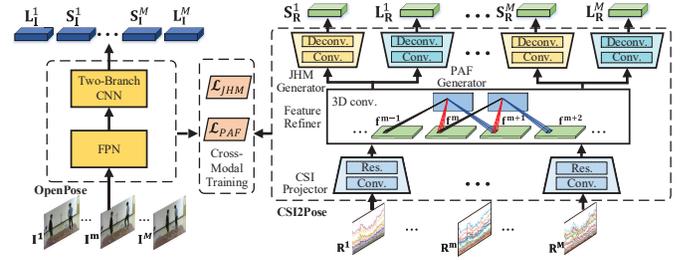}
	\caption{The work flow of semantic feature extraction.}
	\label{fig:network-architecture}
\end{figure}

\textbf{JHM and PAF Extraction.} As shown in Fig.~\ref{fig:network-architecture}, we extract JHM and PAF tensors from  $ \mathcal{I} = \left\lbrace \mathbf{I}^m \right\rbrace_{m=1:M}  $ and $  \mathcal{R} = \left\lbrace \mathbf{R}^m \right\rbrace_{m=1:M}  $.

For video frames $ \mathcal{I}  $, we directly leverage well-known OpenPose network~\cite{cao2019openpose} to output both JHM and PAF tensors in a frame-wise manner. Specifically, OpenPose network $ \mathcal{F_I}(\cdot) $ first adopts a Feature Pyramid Network (FPN)~\cite{lin2017feature} and then utilizes a two-branch CNN to produce visual JHMs and PAFs. After that, we obtain a paired JHM-PAF tensor sequence as
\begin{align}
\left\lbrace (\mathbf{S}^1_{\mathbf{I}}, \mathbf{L}^1_{\mathbf{I}} ) , \cdots, (\mathbf{S}^M_{\mathbf{I}}, \mathbf{L}^M_{\mathbf{I}}) \right\rbrace    =  \mathcal{F_I}(\mathbf{I}^1,\cdots, \mathbf{I}^M),
\end{align}
where $ \mathbf{S}^m_{\mathbf{I}} \in \mathbb{R}^{H \times W \times J} $ and $ \mathbf{L}^m_{\mathbf{I}} \in \mathbb{R}^{H \times W \times 2 \times C } $ are, respectively, JHM and PAF tensors extracted from $ m $-th video frame $ \mathbf{I}^m $.

For RF frames $ \mathcal{R}  $, we develop CSI2Pose network that contains CSI Projector, Feature Refiner, JHM and PAF Generators to extract JHMs and PAFs from time-series Wi-Fi signals.

The CSI projector takes RF frames $ \mathcal{R} $ as input and converts each frame into an image-size tensor that encodes human semantic information under the camera' FOV. For this purpose, the coarse CSI measurements are first tiled to the image size and fed into six 2D convolutional layers to encode wireless features. Then, six residual blocks are used to further embed the wireless features into intermediate feature maps. Specifically, each residual block consists of two 2D convolutional layers and one shortcut connection~\cite{he2016deep}. Each convolutional layer has the kernel size, stride step and padding of 3, 1 and 1, respectively, and is followed by the batch normalization. In addition, the ReLu function is set between the two layers to introduce nonlinearity and the shortcut connection merges the block input and the convolutional result as the final outcome of each residual block~\cite{he2016deep}. Finally, the CSI projector outputs a set of intermediate feature maps $ \mathbf{F} =\left\lbrace \mathbf{f}^1, \cdots , \mathbf{f}^m, \cdots, \mathbf{f}^M  \right\rbrace  $. 

After that, a feature refiner is set up to refine $ \mathbf{F} $. Since human poses usually have high dependencies over the time domain, consecutive RF frames will share a lot of correlations, which can be exploited to boost the effectiveness of the extracted feature maps $ \mathbf{F}  $. To do this, we take the advantages of 3D convolutions~\cite{ji20123d} to learn such temporal correlations. Generally, 3D convolutions are usually exploited to abstract representative features from typical 3D data and have proven their success in video analysis, medical image processing and so on. Analogous to video data, the set of feature maps $ \mathbf{F} $ can be viewed as a 3D cube that is stacked by a sequence of image-size tensors, making 3D convolutions suitable for our refinement task. In this way, we employ two layers of 3D convolutions on adjacent feature maps to enhance human semantic features along the temporal dimension. Finally, our feature refiner generates a set of $ M $ refined feature maps $ \tilde{\mathbf{F}} =\left\lbrace \tilde{\mathbf{f}^1} , \cdots ,  \tilde{\mathbf{f}^m}, \cdots, \tilde{\mathbf{f}^M} \right\rbrace $.

In the following two generators, we transform each refined feature map $ \tilde{\mathbf{f}^m} $ into JHM and PAF spaces, respectively. To do this, $ \tilde{\mathbf{f}^m} $ is first processed by two layers of 2D convolutions. Due to downsampling operations, i.e., convolutions, the width and height of intermediate features are smaller than those of JHMs and PAFs. Thus, it is necessary to upsample these learned features into high-resolution ones. Inspired by Fully Convolutional Network (FCN)~\cite{long2015fully}, which exploits deconvolutions for pixel-level image semantic segmentation, we utilize two layers of deconvolutions with the bilinear mode to transform intermediate feature maps into desirable image-size JHM and PAF tensors. As JHMs and PAFs convey different semantic information, we build two FCNs for the JHM generator and the PAF generator, respectively, to produce a JHM tensor $ \mathbf{S}^m_{\mathbf{R}} \in \mathbb{R}^{H \times W \times C} $ and a PAF tensor $ \mathbf{L}^m_{\mathbf{R}} \in \mathbb{R}^{H \times W \times C \times 2} $ for $ \tilde{\mathbf{f}^m} $.

After all, given the RF frames $ \mathcal{R} $, our CSI2Pose network outputs a paired JHM-PAF sequence as
\begin{align}
\left\lbrace ( \mathbf{S}^1_{\mathbf{R}}, \mathbf{L}^1_{\mathbf{R}} ), \cdots, ( \mathbf{S}^M_{\mathbf{R}}, \mathbf{L}^M_{\mathbf{R}} ) \right\rbrace= \mathcal{F_R} (\mathbf{R}^1,\cdots, \mathbf{R}^M),
\end{align}
where $ \mathcal{F_R}(\cdot) $ are trainable parameters of CSI2Pose network. 


\textbf{Cross-Modal Training.} We train CSI2Pose network with automatic semantic labels for avoiding laborious manual annotation. To achieve this, there are basically marker-based and maker-less approaches. The marker-based approaches attach special markers on the subject body and track these markers using dedicated motion capturing systems, such as VICON~\cite{jiang2020towards}. Though producing high-precision keypoint estimations, these approaches require lots of calibrated cameras to be deployed around the target area. In contrast, the mark-less approaches leverage computer vision techniques, such as OpenPose~\cite{cao2019openpose}, to output reliable body keypoint annotations without dedicated camera systems and cumbersome deployment~\cite{zhao2018rf}. For these reasons, we adopt the marker-less approach and train our CSI2Pose network with the supervision of OpenPose network.

\begin{figure}
	\centering
	\subfigure[PDFs of element values.]{
		\includegraphics[width=0.475\linewidth]{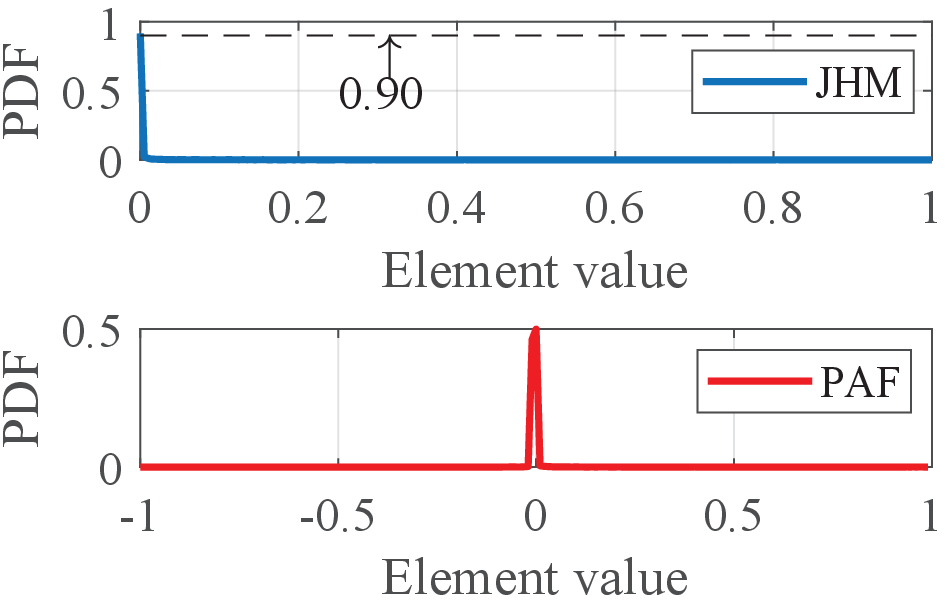}}
	\label{1c}
	\subfigure[Elements-wise loss weights.]{
		\includegraphics[width=0.475\linewidth]{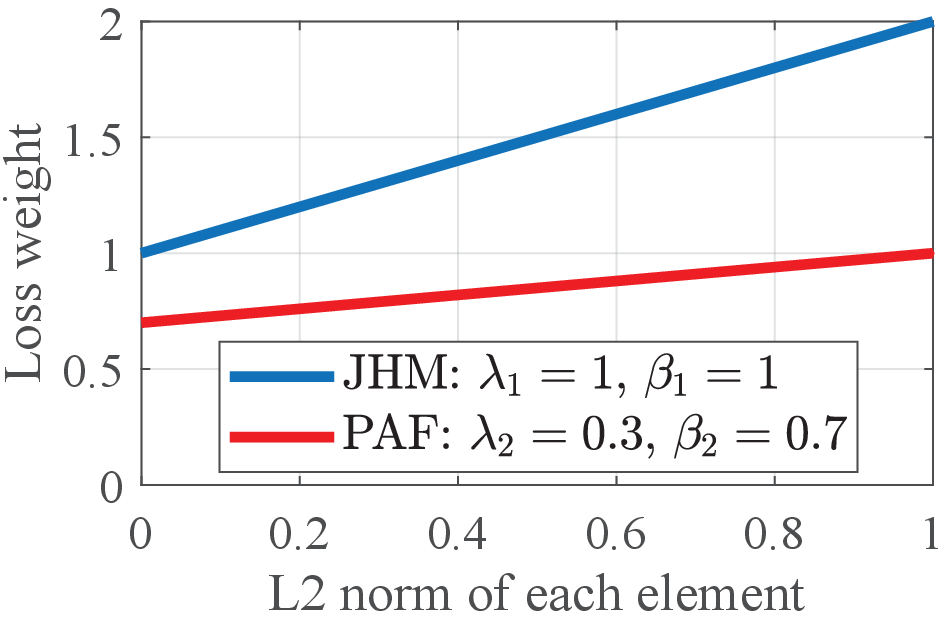}}
	\label{2c}\\
	\caption{PDFs and training loss weights for visual JHMs and PAFs.}
	\label{fig:visualjhmpaf} 
\end{figure}

Specifically, given a training set of video-RF frame sequences $ \left\lbrace (\mathcal{I}^y,\mathcal{R}^y)  \right\rbrace_{y=1:Y}  $, we first input $ \left\lbrace \mathcal{I}^y \right\rbrace_{y=1:Y}  $ into OpenPose network $ \mathcal{F_{I}} (\cdot) $ and obtain the visual feature set $ \left\lbrace  (\mathbf{S}^{ym}_{\mathbf{I}}, \mathbf{L}^{ym}_{\mathbf{I}}) \right\rbrace_{y=1:Y, m=1:M}  $. In this way, the training objective is to minimize the discrepancy between outputs of CSI2Pose and OpenPose networks:
\begin{align}\label{eq: training object}
\mathop{\min} \limits_{\mathcal{F_{R}}} \frac{1}{Y} \sum_{ym} \mathcal{L}_{JHM} \left( \mathbf{S}^{ym}_{\mathbf{I}}, \mathbf{S}^{ym}_{\mathbf{R}} \right) + \mathcal{L}_{PAF} \left( \mathbf{L}^{ym}_{\mathbf{I}}, \mathbf{L}^{ym}_{\mathbf{R}} \right) ,
\end{align}
where $ \mathcal{L}_{JHM} (\cdot,\cdot) $ and $ \mathcal{L}_{PAF} (\cdot,\cdot) $ are loss functions for JHM and PAF features and can be further expressed as
\begin{align}
\mathcal{L}_{JHM} \left( \mathbf{S}_{\mathbf{I}}, \mathbf{S}_{\mathbf{R}} \right)= \sum_{j} \sum_{hw} \alpha^j_{hw}  || \mathbf{s}^j_{\mathbf{I}} (h,w) - \mathbf{s}^j_{\mathbf{R}} (h,w)||^2_2, \label{eq:loss jhm} \\
\mathcal{L}_{PAF} \left( \mathbf{L}_{\mathbf{I}}, \mathbf{L}_{\mathbf{R}} \right)= \sum_{c} \sum_{hw} \alpha^c_{hw}  || \mathbf{l}^c_{\mathbf{I}} (h,w) - \mathbf{l}^c_{\mathbf{R}} (h,w) ||^2_2. \label{eq:loss paf}
\end{align}
In Eq.~\eqref{eq:loss jhm} and Eq.~\eqref{eq:loss paf}, $ \alpha^j_{hw} $ and $ \alpha^c_{hw} $ are element-wise weights for JHM and PAF tensors, respectively, and need careful configurations for our cross-modal training. For this purpose, we plot the probability density functions (PDFs) of element values for visual JHMs and PAFs in Fig.~\ref{fig:visualjhmpaf} (a). As the figure shows, we can observe that up to 90\% of elements in JHMs have nearly zero values and only less than 10\% indicate useful body keypoints. The similar observation can be found in PAFs. Hence, for giving more attention on informative human semantic features, we let both $ \alpha^j_{hw} $ and $ \alpha^c_{hw} $ to be linearly increasing with the L2 norm of $ (h,w) $-th element in the ground-truth labels $ \mathbf{S}_{\mathbf{I}} $ and $ \mathbf{L}_{\mathbf{I}} $. In this way, we define $ \alpha^j_{hw} = \lambda_1  ||\mathbf{s}^j_{\mathbf{I}}(h,w)||_2 + \beta_1 $ and $ \alpha^c_{hw} = \lambda_2  ||\mathbf{l}^c_{\mathbf{I}}(h,w)||_2 + \beta_2 $. In our system, we empirically set $\lambda_1=1$, $\beta_1=1$,  $\lambda_2=0.3$ and $\beta_2=0.7$. The corresponding loss weights for JHMs and PAFs are plotted in Fig.~\ref{fig:visualjhmpaf} (b).

\subsection{Forgery Detection and Localization}

\begin{figure}
	\centering
	\includegraphics[width=0.95\linewidth]{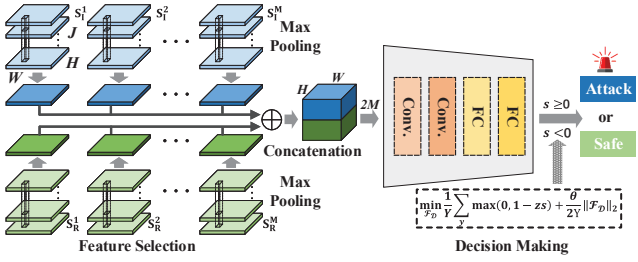}
	\caption{The architecture of our detection network.}
	\label{fig:detectionmetwork}
\end{figure}


\textbf{Forgery Detection.} We formulate the forgery detection as a binary classification task, which maps $ \left\lbrace( \mathbf{S}^m_{\mathbf{I}} , \mathbf{S}^m_{\mathbf{R}} ) \right\rbrace_{m=1:M}  $ into a decision space $  \mathcal{Z} = \left\lbrace -1, 1 \right\rbrace $. Therein, we let $ z = 1 $ to represent a positive event that indicates forgery attacks in $ \mathcal{I} $. We design a lightweight detection network $ \mathcal{F_D}(\cdot) $ that has two stages for feature selection and decision making as shown in Fig.~\ref{fig:detectionmetwork}. 

At the first stage, our detection network has two branches to squeeze visual and wireless JHM tensors into single-channel feature planes for reducing computational complexity. As depicted in Fig.~\ref{fig:visualjhmpaf} (a), positive elements are sparsely distributed in JHMs, suggesting that only a tiny portion of pixels in each channel are occupied by body keypoints. Hence, it is desired to compact such sparse tensors before convolutional operations. To do this, for $ (h,w) $-th location at one JHM tensor, we perform max pooling along the channel dimension, which selects the maximal joint response to represent the body presence at that location. In this way, the input size of the first convolutional layer can be reduced by $ J=14 $ times from $ 2 \times H \times W \times J \times M $ to $2 \times H \times W \times M $. Then, we stack compacted visual and wireless JHMs along the temporal dimension, respectively, and thereafter concatenate two results into one 3D feature tensor with a size of $  H \times W \times 2M $.

At the second stage, our detection network first employs two convolutional layers, consisting of 64 and 32 convolution kernels with the size and stride of 5 and 2, respectively, to abstract forgery traces. Next, it adopts two fully connected (FC) layers with 672 and 256 neurons, where the ReLu function is used for activation at each layer. Finally, the Tanh function is leveraged to yield a scalar value $ s \in [-1, 1] $ that stands for the likelihood of attack occurrence in $ \mathcal{I} $. To this end, our detection network can make a decision $ z'=1 $ if $ s>0 $; otherwise, $ z' = -1 $.
 
In the training phase, given a set of visual-wireless JHM pairs  $ \left\lbrace( \mathbf{S}^{ym}_{\mathbf{I}} , \mathbf{S}^{ym}_{\mathbf{R}} ) \right\rbrace_{y=1:Y, m=1:M}  $ and its label set $ \left\lbrace z^y \right\rbrace_{y=1:Y}  $, we train our detection network using the hinge loss to force the network to output high-confidence predictions. Specifically, the hinge loss uses the operation $ \max \left( 0, 1-zs \right)  $ to compute the distance between $ z $ and $ s $. This operation not only makes the network to output a right label, i.e.,  $ s \le 0 $ or $ s > 0 $, but also forces it to produce a high-confidence outcome, i.e., driving $ s $ to approach $ z $ as close as possible. In this way, our detection network can learn a clear decision boundary between samples of two classes and thus will have a high detection accuracy in the testing phase. Moreover, a L2 norm regularization is adopted for alleviating over-fitting. Therefore, the final training process can be formulated as the following optimization problem:
\begin{align}\label{eq: training detection network}
\mathop{\min} \limits_{\mathcal{F_{D}}} \frac{1}{Y} \sum_y \max \left(0,1-zs \right) + \frac{\theta}{2Y} ||\mathcal{F_D}||_2 , 
\end{align}
where $ \theta $ is a hyperparameter.

\begin{figure}
	\centering
	\includegraphics[width=0.95\linewidth]{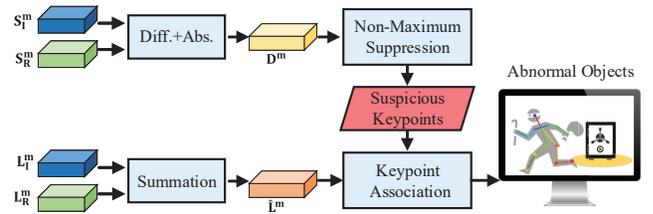}
	\caption{The work flow of video forgery localization.}
	\label{fig:detectionlocalization}
\end{figure}

\textbf{Forgery Localization.} Once $ \mathcal{I} $ is detected as a forged GOP, our system goes deep into each video frame to localize forgery traces. Given JHMs and PAFs, a straightforward approach is to first localize all possible human instances via associating the detected body keypoints in each frame using body part association in~\cite{cao2019openpose} and then find mismatched human instances in a video-RF frame pair in a person-by-person manner. However, such approach is cumbersome and computationally inefficient. In contrast, Secure-Pose focuses on localizing abnormal human objects only in forged regions, while ignoring unnecessary legal ones. This attention scheme will make our system more computationally efficient than the algorithm in~\cite{cao2019openpose} in the task of video forgery localization.  

As shown in Fig.~\ref{fig:detectionlocalization}, for each suspicious video frame $ \mathbf{I}^m $, the first step is to select keypoint candidates of abnormal objects based on the estimated visual and wireless JHMs $ \mathbf{S}^m_{\mathbf{I}} $ and $ \mathbf{S}^m_{\mathbf{R}} $. For this purpose, we compute the absolute values of element-wise differences between $ \mathbf{S}^m_{\mathbf{I}} $ and $ \mathbf{S}^m_{\mathbf{R}} $ as
\begin{align}\label{eq:jhm residual}
\mathbf{D}^m = \left|  \mathbf{S}^m_{\mathbf{I}} - \mathbf{S}^m_{\mathbf{R}} \right| ,
\end{align}
where $ \mathbf{D}^m  \in \mathbb{R}^{H\times W \times J} $. The rationale behind Eq.~\eqref{eq:jhm residual} is that when video forgery attacks appear, body poses in the video frame $ \mathbf{I}^m $ is modified and the visual JHMs $ \mathbf{S}^m_{\mathbf{I}} $ will change accordingly, resulting in large differences between $ \mathbf{S}^m_{\mathbf{I}} $ and $ \mathbf{S}^m_{\mathbf{R}} $ in the forged regions and near-zero values in the genuine areas. Afterwards, our system performs non-maximum suppression on $ \mathbf{D}^m $ to select a set of suspicious keypoints $ \mathcal{K}^m $:
\begin{align}
\mathcal{K}^m =\left\lbrace \mathbf{k}^m_{j,n} \in \mathbb{R}^2 : \text{for} j \in \left\lbrace 1,\cdots ,J  \right\rbrace, n \in \left\lbrace 1, \cdots, N_j \right\rbrace  \right\rbrace,
\end{align}
where $ N_j $ denotes the number of candidates of $ j $-th keypoint and $ \mathbf{k}^m_{j,n}  $ is the location of $ n $-th candidate of $ j $-th keypoint. 

The second step is to associate the suspicious keypoints $ \mathcal{K}^m $ to form abnormal objects using PAFs  $ \mathbf{L}^m_{\mathbf{I}} $ and $ \mathbf{L}^m_{\mathbf{R}} $. In this step, our system first merges  $ \mathbf{L}^m_{\mathbf{I}} $ and $ \mathbf{L}^m_{\mathbf{R}} $ into a combined PAF tensor $ \hat{\mathbf{L}}^m \in \mathbb{R}^{H\times W \times 2 \times C} $ before the keypoint association as
\begin{align}\label{eq: combined paf}
\hat{\mathbf{L}}^m = \mathbf{L}^m_{\mathbf{I}} + \mathbf{L}^m_{\mathbf{R}}.
\end{align}
The reason of this operation is that if only $ \mathbf{L}^m_{\mathbf{I}} $ or $ \mathbf{L}^m_{\mathbf{R}} $ is used, the information about abnormal object in forged regions may be missing. However, their summation result can easily reserve such information. Then, abnormal human poses are determined by a set of connected body keypoints, and such best connection states $ \mathcal{E}^m  $ between keypoints in $ \mathcal{K}^m $ can be obtained by 
\begin{align}
\mathcal{E}^m = \mathcal{F_A} \left( \mathcal{K}^m, \hat{\mathbf{L}}^m \right), 
\end{align}
where $  \mathcal{F_A} (\cdot,\cdot) $ represents the association function used in~\cite{cao2019openpose}. Moreover, each element in $ \mathcal{E}^m  $ is a binary variable $ E_{k_1,k_2} \in \left\lbrace 0,1 \right\rbrace  $, which indicates the connection state between $ k_1 $-th and $ k_2 $-th keypoints in $ \mathcal{K}^m $. Toward this end, Secure-Pose can produce the body poses of abnormal objects in forged regions.

\begin{figure}
	\centering
	\includegraphics[width=0.95\linewidth]{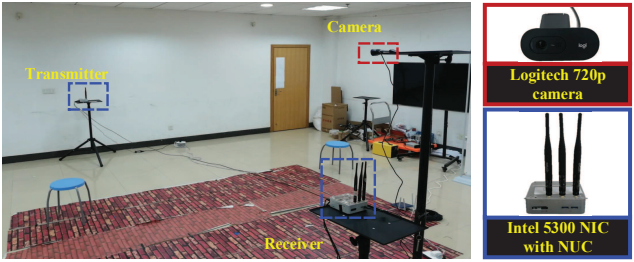}
	\caption{Experimental platform using one camera and two Intel 5300 NICs.}
	\label{fig:experimentalsetup}
\end{figure}

\section{Implementation and Evaluation}\label{sec:implementation}

\subsection{Implementation}

As shown in Fig.~\ref{fig:experimentalsetup}, we implement Secure-Pose using one Logitech 720p camera and two Intel 5300 NICs. Specifically, NICs are mounted on two Intel D54350WYKH NUCs, each of which has a quad-core Core i5 processor @1.30 GHz as well as an 8 GB RAM. Each NIC is equipped with three omnidirectional VERT2450 antennas, uniformly spaced with a distance of 2.6 cm, to act as a transmitter or receiver. Moreover, we place two NICs at a height of one meter and separate them with each other at a distance of about six meters. For a good FOV, we mount the camera with a height of 2 m and place it on the receiver side. To generate video signals, the camera is connected to the receiving NUC and set to output 1280$\times$720p RGB videos with a FPS of 7.5 Hz and a GOP size of 12. To sample CSI signals, we use the 802.11n CSI Tool~\cite{Halperin2011csitool} to control two NICs to communicate with a bandwidth of 20 MHz centering in the 5.6 GHz Wi-Fi band. In this condition, we record CSI measurements of 30 subcarriers with a sampling rate of 100 Hz.

\subsection{Evaluation Methodology}
 
\textbf{Data Collection.} We collect concurrent camera and Wi-Fi data in different environments including two office rooms and a corridor in our campus. In each environment, the number of people under the camera's FOV is set to vary from zero to four. In each setting, we allow the involving participants to behave casually for about one minute, and the above trial is repeated for six times. During each trial, each participating individual can freely perform static motions, such as sitting on a chair and standing still, and dynamic motions, such as walking and waving hands. Moreover, all involving participants can have different body motions at the same time. In this way, the collected Wi-Fi signals would have influences from both static and dynamic body motions. Finally, we collect multi-modal data of about half an hour in each environment. 

\textbf{Dataset.} After data collection, we synchronize CSI and video data using their timestamps, denoise the CSI data, and obtain more than 11K synchronized video-RF fame pairs in each environment. With these samples, we perform playback attacks by randomly replacing one pristine GOP with another one. To launch tampering attacks, we leverage Faster-RCNN~\cite{ren2017faster} to detect and crop partial human objects out and then replace them with corresponding blank background segments. The left GOPs with authentic video frames are considered as negative samples. Finally, we make a labeled dataset for each environment and thus obtain a total of three different datasets, each of which has the equal numbers of positive and negative samples. Particularly, the 70\% of samples of each dataset are used for training and validation and the left 30\% for testing. Based on three labeled datasets, we train and test our system on each of them, respectively, and average their metrics as a final output. 

\begin{figure}
	\centering
	\includegraphics[width=0.87\linewidth]{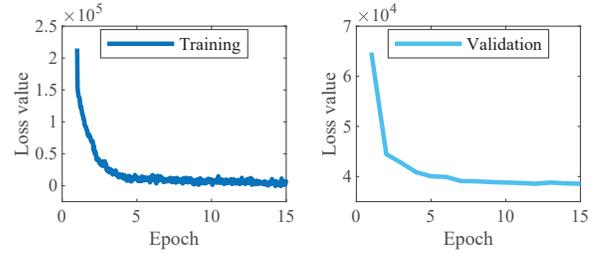}
	\caption{The training and validation losses of CSI2Pose network.}
	\label{fig:trainingloss}
\end{figure}

\textbf{Training Details.} We train our CSI2Pose and detection networks with the batch size of 1 and 32, respectively, on the PyTorch framework. A RMSprop optimizer with a weight decay of 1e-6 and a momentum of 0.9 is adopted. Since the loss functions $ \mathcal{L}_{JHM} $ and $ \mathcal{L}_{PAF} $ represent the sums of the differences between the visual and wireless semantic features over all elements, the value of CSI2Pose's training loss would be very large, which leads to large-value gradients and makes the training process unstable. For effective training, we scale down such gradients into a reasonable range by empirically setting the initiate learning rate to be 1e-5 for CSI2Pose network. Additionally, we set that of the detection network to be 1e-4. Moreover, the learning rates will be multiplied by a decay factor 0.3 after every 5 epochs. Finally, we train CSI2Pose network with 15 epochs and the detection network with 5 epochs. We depict the training and validation losses of CSI2Pose network in Fig.~\ref{fig:trainingloss}. As the figure shows, the above settings can result in effective training of our network.

\textbf{Evaluation Metrics.} We leverage the accuracy, true positive rate (TPR) and false positive rate (FPR) to evaluate the detection performance of our system and use percentage of correct keypoint (PCK) to measure its localization performance.
\begin{itemize}
	\item \textbf{Accuracy.} It is the ratio of the correctly-detected samples to the total number of samples.
	\item \textbf{TPR.} It is the ratio of the successfully-detected positive samples to the total number of positive samples. 
	\item \textbf{FPR.} It is the ratio of the mistakenly-recognized negative samples to the total number of negative samples.
	\item \textbf{PCK.} It is the ratio that the normalized distance from prediction $ \mathbf{x}^{j}_{p} $ to ground-truth $ \mathbf{y}^{j}_p $ of $ j $-th keypoint of $ p $-th person is less than $ \rho  $: $ \text{PCK}_j@\rho = \frac{1}{P} \sum_{p} \mathbbm{1} \left\lbrace \frac{||\mathbf{x}^{j}_{p}-\mathbf{y}^{j}_p||_2}{b^p} \le \rho \right\rbrace  $, where $ 0< \rho < 1 $, $ P $ is the number of people and $ b^p $ is the diagonal length of $ p $-th person's bounding box. Moreover, the mean PCK over all keypoints is denoted as $ \text{mPCK}$.
\end{itemize}

\begin{figure}
	\centering
	\includegraphics[width=0.87\linewidth]{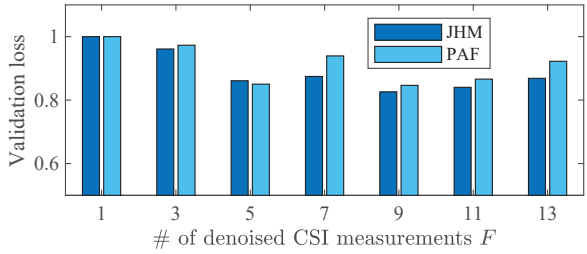}
	\caption{Validation loss using different $ F $ in each RF frame.}
	\label{fig:parameterdetermination}
\end{figure}

\subsection{Experimental Results}

\textbf{Parameter Determination.} The first step of our experiment is to decide the number of denoised CSI measurements $ F $ assigned to each RF frame. To do this, we compute the normalized validation losses of CSI2Pose network defined in Eq.\eqref{eq:loss jhm} and Eq.\eqref{eq:loss paf}. As shown in Fig.~\ref{fig:parameterdetermination}, the JHM loss generally decreases with $ F $ at the beginning. This is because that the larger $ F $ is, the finer-grained information is offered for each video frame, consequently yielding the higher regression performance. However, when $ F $ exceeds 9, it becomes steady and even increases slightly. The same trend can be found in the PAF loss. Based on the above observations, we assign nine denoised CSI measurements in each RF frame.

\begin{table}[]
	\centering
	\caption{Detection Results in Different Types of Attacks.}\label{tab:detection results}
	\begin{tabular}{|c|c|c|c|c|}
		\hline
		& \textbf{TPR} & \textbf{FPR} & \textbf{Acc.} & \textbf{AUROC} \\ \hline
		\textbf{Playback} & 99.3\%       & 1.0\%        & 99.2\%        & 0.999          \\ \hline
		\textbf{Tampering} & 99.1\%       & 2.6\%        & 98.3\%        & 0.995          \\ \hline
		\textbf{Overall}     & 99.2\%       & 1.8\%        & 98.7\%        & 0.997          \\ \hline
	\end{tabular}
\end{table}

\textbf{Forgery Detection Performance.} Based on the determined $ F $, we present the detection performance of Secure-Pose in our experiment. To do this, we compute TPR, FPR and accuracy under playback and tampering attacks. As shown in Table~\ref{tab:detection results}, Secure-Pose achieves better performance under playback attacks. This is due to that playback attacks would probably mismatch all human objects in authentic video streams and thus incur large differences between authentic and forged frames, making our system easier to detect such forgeries. Despite that, our system still has very close performance
in two attack scenarios. To sum up, Secure-Pose achieves an average detection accuracy of 98.7\%. Specifically, it can successfully detect 99.2\% of forged video streams and correctly recognize 98.2\% authentic ones. After that, we also calculate area under the receiver operating characteristic curve (AUROC) to comprehensively indicate its discrimination capability. As reported in Table~\ref{tab:detection results}, our system obtains a high overall AUROC of 0.997, which is close to the ideal case. The above results demonstrate the high effectiveness of Secure-Pose in detecting various forgery attacks in video streams.

\begin{figure}
	\centering
	\includegraphics[width=0.88\linewidth]{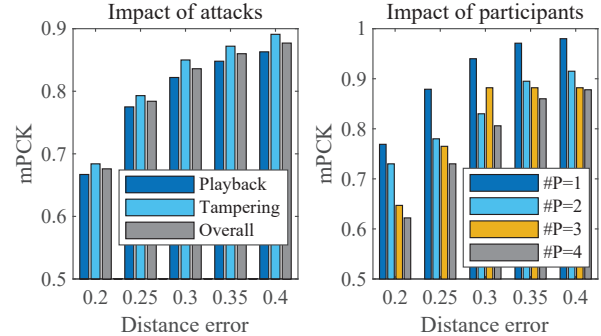}
	\caption{Localization performance under different attacks and different numbers of participants(\#P).}
	\label{fig:forgerylocalizationperformance}
\end{figure}

\begin{table}[]
	\centering
	\caption{Detection Performance in Different Environments.}\label{fig:detectionperformanceinenvironments}
	\begin{tabular}{|c|c|c|c|c|}
		\hline
		& \textbf{TPR} & \textbf{FPR} & \textbf{Acc.}  \\ \hline
		\textbf{Office one} & 99.2\%       & 1.9\%        & 98.6\%           \\ \hline
		\textbf{Office two} & 99.3\%       & 2.3\%        & 98.5\%          \\ \hline
		\textbf{Corridor}     & 99.2\%       & 1.4\%        & 98.9\%        \\ \hline
	\end{tabular}
\end{table}

Moreover, we demonstrate Secure-Pose's detection performance in different environments. As depicted in Table~\ref{fig:detectionperformanceinenvironments}, our system has nearly the same detection performance in the three different settings, which suggests the robustness of our system against ambient environments. In addition, environmental noise in Wi-Fi signals would have an impact on the system's performance. To deal with such noise, Secure-Pose first adopts a customized low-pass Butterworth filter to remove high-frequency noise from ambient environments and takes a sequence of successive signal signatures for extracting reliable human semantic features. As shown in Table~\ref{tab:detection results}, Secure-Pose obtains an overall detection accuracy of 98.7\%, suggesting that it has the ability to eliminate real-world environmental noise.

\begin{figure}
	\centering
	\includegraphics[width=0.87\linewidth]{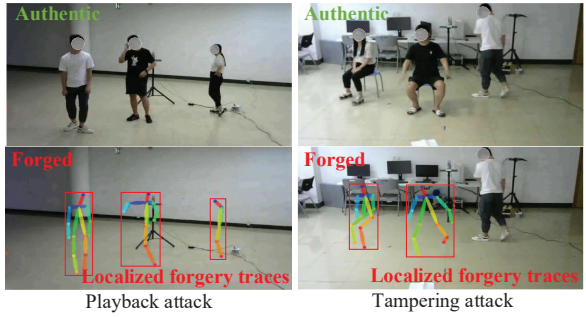}
	\caption{Instances of localized forgery traces under different attacks.}
	\label{fig:instancelocalization}
\end{figure}

\textbf{Forgery Localization Performance.} Then, we illustrate the forgery localization performance of our system. For this purpose, we first show the mean PCKs of pose estimation of abnormal objects under different forgery attacks. As depicted in Fig.~\ref{fig:forgerylocalizationperformance}, our system yields high mean PCKs at low normalized distance errors. For instance, there are about 80\% estimated locations of abnormal keypoints that are within 25\% of diagonal lengths of person bounding boxes. Moreover, it also can be observed that our system always has a higher mean PCK under tampering attacks. The reason is that compared with playback attacks, tampering attacks have a smaller modification on $ \mathbf{D}^m $ in Eq.~\eqref{eq:jhm residual} and $ \hat{\mathbf{L}}^m $ in Eq.~\eqref{eq: combined paf}, which contributes to a higher accuracy on keypoint localization and association for abnormal objects. Afterwards, we investigate the impact of the number of participants on localization performance. As demonstrated in Fig.~\ref{fig:forgerylocalizationperformance}, we can observe that basically the fewer the participants are, the higher localization performance is achieved. This is due to that the human body has a great impact on Wi-Fi signals. The more bodies are present under the camera's FOV, the more dynamic CFRs are incurred in CSI measurements, rendering our system more difficult to recover body keypoints.
	
In addition, we showcase localized forgery traces under playback and tampering attacks, respectively, in Fig.~\ref{fig:instancelocalization}. In the playback attack, we replace the authentic frame that has three persons with a blank background. In this case, the absence of three persons is successfully detected by our system, and then their associated body keypoints and bounding boxes are accurately inferred and presented on the forged frame. In the tampering attack, two of three persons are erased from the authentic frame. In this case, our system correctly localizes forged areas and estimates the body poses of the erased persons. The above results suggest the effectiveness of Secure-Pose in localizing forgery traces in video streams.

\begin{figure}
	\centering
	\includegraphics[width=0.9\linewidth]{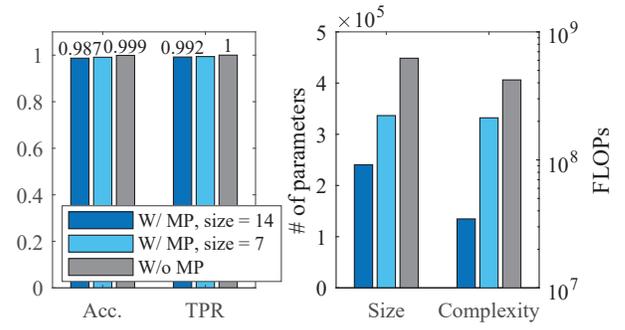}
	\caption{The impact of max pooling (MP) on our detection network.}
	\label{fig:detectionwithwithoutpooling}
\end{figure}

\textbf{Effectiveness of Detection Network.} Next, we explore the impact of max pooling on the proposed detection network. For this purpose, we build two baseline models. In particular, one baseline has a pool size of 7, meaning that two local maximums are selected at each position, and the other skips the stage of feature selection and directly concatenates both visual and wireless JHM features as its input. We compare their detection performance in terms of accuracy and TPR, and we also report their network size and computational complexity using the number of network parameters and floating point operations (FLOPs), respectively. As shown in Fig.~\ref{fig:detectionwithwithoutpooling}, compared with the baseline without feature selection, max pooling with a pool size of 14 causes a tiny accuracy decrease of 1.2\% and a subtle TPR decrease of 0.8\%, while it contributes to a significant size reduction by about 50\% and the computational complexity one order lower. The reason is that JHMs are very sparse features in the image space and the maximum over all channels can provide sufficient information about body presence at each position. However, when we further decrease the pool size to 7, i.e., doubling the input features of convolutional layers, only marginal gains are obtained in detection performance. Moreover, the smaller pool size leads to a size increase of about 40\% and a complexity increase of about six times. The above observations verify that max pooling on 14 channels can effectively reduce both the network size and computational complexity of our detection network, while having little impact on its detection performance.

\begin{figure}
	\centering
	\includegraphics[width=0.92\linewidth]{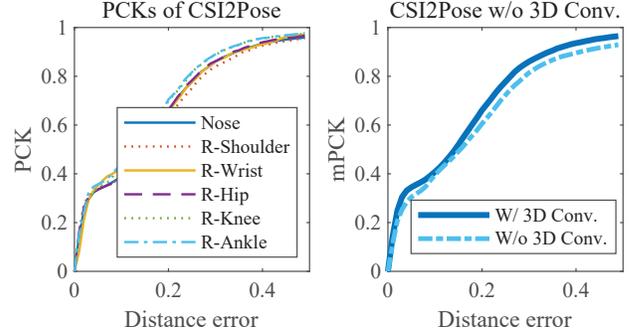}
	\caption{PCKs of CSI2Pose network. Right-side keypoints are present.}
	\label{fig:keypointpck}
\end{figure}

\textbf{Effectiveness of CSI2Pose Network.} Then, we show the performance of CSI2Pose network. To measure its ability of pose estimation, we directly leverage body association method~\cite{cao2019openpose} on CSI2Pose network's outputs. As depicted in Fig.~\ref{fig:keypointpck}, our system yields similar results for all body keypoints. However, the keypoints of knees and ankles generally obtain slightly higher PCKs. This may be due to that the knee and ankle parts basically have less opportunities to be occluded when compared with other keypoints, such as shoulders and hips, making OpenPose network to produce higher-quality ground-truth labels at the training stage. Moreover, we illustrate the merits of 3D convolutions adopted in our CSI2Pose network. To do this, we show its mean PCKs with and without 3D convolutions. As Fig.~\ref{fig:keypointpck} depicts, the adoption of 3D convolutions makes CSI2Pose network to generate higher mean PCKs. The results suggest the effectiveness of 3D convolutions in boosting the performance of our CSI2Pose network.
  
\begin{table}[]
	\centering
	\caption{Test Runtime of Our System.}\label{tab:runtime}
	\begin{tabular}{|c|c|c|c|c|c|}
		\hline
		& \textbf{OpenPose} & \textbf{CSI2Pose} & \textbf{Detect} &  \textbf{Localize} & \textbf{Total} \\ \hline
		\textbf{Runtime } &     551.0 ms       &            11.1 ms    &   0.8 ms &     0.1 ms       &                     563.0   ms         \\ \hline
	\end{tabular}
\end{table}

\textbf{Runtime Analysis.} Finally, to verify the real-time ability of our system, we run it on a laptop with one NVIDIA Tesla P40 GPU and report the runtime of each component for processing one testing GOP. As shown in Table~\ref{tab:runtime}, compared with OpenPose and CSI2Pose networks, the component of localization has a negligible test runtime due to its low computation complexity. Moreover, OpenPose incurs a much higher runtime than CSI2Pose. The reason is that OpenPose processes each GOP in a frame-by-frame manner, while CSI2Pose takes the whole sequence as one input, which contributes to high GPU utilization. Despite that, our system has a total runtime of about 0.5 s for each GOP. Considering that the camera generates videos with a FPS of 7.5 Hz and a GOP size of 12 in our experiment, the duration of each GOP can be computed as $ 1/ 7.5 \times 12 = 1.6 \, \text{s}$. Under these conditions, our system can finish the forgery detection and localization on the current GOP before the reception of the next one, which suggests the real-time ability of our system to detect and localize video forgeries.

\section{Conclusion}\label{sec:conclusion}
This paper presents Secure-Pose, a novel cross-modal system that effectively detects and localizes forgery traces in live surveillance videos using ambient Wi-Fi signals. We observe that the coexisting camera and Wi-Fi signals contain common human semantic information and the presence of video forgery attacks will decouple such cross-modal information correspondence. Our system effectively extracts sparse human semantic features from synchronized camera and Wi-Fi signals and efficiently detects and localizes forgery traces in video streams. We implement our system using one Logitech 720p camera and two Intel 5300 NICs and evaluate it in different indoor environments. The evaluation results demonstrate that Secure-Pose achieves a high detection accuracy of 98.7\% and accurately localizes abnormal objects under both playback and tampering attacks.

\bibliographystyle{IEEEtran}
\bibliography{IEEEabrv,./Forgerydetection}

\begin{thebibliography}{10}
\providecommand{\url}[1]{#1}
\csname url@samestyle\endcsname
\providecommand{\newblock}{\relax}
\providecommand{\bibinfo}[2]{#2}
\providecommand{\BIBentrySTDinterwordspacing}{\spaceskip=0pt\relax}
\providecommand{\BIBentryALTinterwordstretchfactor}{4}
\providecommand{\BIBentryALTinterwordspacing}{\spaceskip=\fontdimen2\font plus
\BIBentryALTinterwordstretchfactor\fontdimen3\font minus
  \fontdimen4\font\relax}
\providecommand{\BIBforeignlanguage}[2]{{%
\expandafter\ifx\csname l@#1\endcsname\relax
\typeout{** WARNING: IEEEtran.bst: No hyphenation pattern has been}%
\typeout{** loaded for the language `#1'. Using the pattern for}%
\typeout{** the default language instead.}%
\else
\language=\csname l@#1\endcsname
\fi
#2}}
\providecommand{\BIBdecl}{\relax}
\BIBdecl

\bibitem{yong2021towards}
Y.~{Huang}, X.~{Li}, W.~{Wang}, T.~{Jiang}, and Q.~{Zhang}, ``Towards
  cross-modal forgery detection and localization on live surveillance videos,''
  in \emph{Proc.~IEEE INFOCOM}, 2021.

\bibitem{liu2013intelligent}
H.~Liu, S.~Chen, and N.~Kubota, ``Intelligent video systems and analytics: A
  survey,'' \emph{IEEE Trans. Ind. Informat.}, vol.~9, no.~3, pp. 1222--1233,
  2013.

\bibitem{zhong2017multi}
C.~Zhong, X.~Jiang, F.~Qu, and Z.~Zhang, ``Multi-antenna wireless legitimate
  surveillance systems: Design and performance analysis,'' \emph{IEEE Trans.
  Wireless Commun.}, vol.~16, no.~7, pp. 4585--4599, 2017.

\bibitem{exploiting2013}
\BIBentryALTinterwordspacing
``{Exploiting Network Surveillance Cameras Like a Hollywood Hacker},'' craig
  Heffners, 2013. [Online]. Available:
  \url{https://www.youtube.com/watch?v=B8DjTcANBx0}
\BIBentrySTDinterwordspacing

\bibitem{looping2015}
\BIBentryALTinterwordspacing
``{Looping Surveillance Cameras through Live Editing},'' van Albert and Bank,
  2015. [Online]. Available: \url{https://www.youtube.com/watch?v=RoOqznZUClI}
\BIBentrySTDinterwordspacing

\bibitem{fayyaz2020improved}
M.~A. Fayyaz, A.~Anjum, S.~Ziauddin, A.~Khan, and A.~Sarfaraz, ``An improved
  surveillance video forgery detection technique using sensor pattern noise and
  correlation of noise residues,'' \emph{Multimedia Tools and Applications},
  vol.~79, no.~9, pp. 5767--5788, 2020.

\bibitem{yang2016using}
J.~Yang, T.~Huang, and L.~Su, ``Using similarity analysis to detect frame
  duplication forgery in videos,'' \emph{Multimedia Tools and Applications},
  vol.~75, no.~4, pp. 1793--1811, 2016.

\bibitem{chen2015automatic}
S.~Chen, S.~Tan, B.~Li, and J.~Huang, ``Automatic detection of object-based
  forgery in advanced video,'' \emph{IEEE Trans. Circuits Syst. Video
  Technol.}, vol.~26, no.~11, pp. 2138--2151, 2015.

\bibitem{ulutas2017frame}
G.~Ulutas, B.~Ustubioglu, M.~Ulutas, and V.~Nabiyev, ``Frame
  duplication/mirroring detection method with binary features,'' \emph{IET
  Image Processing}, vol.~11, no.~5, pp. 333--342, 2017.

\bibitem{wang2007exposing}
W.~Wang and H.~Farid, ``Exposing digital forgeries in video by detecting
  duplication,'' in \emph{Proc.~ACM MM\&Sec}, 2007, pp. 35--42.

\bibitem{lakshmanan2019surfi}
N.~Lakshmanan, I.~Bang, M.~S. Kang, J.~Han, and J.~T. Lee, ``Surfi: detecting
  surveillance camera looping attacks with {Wi-Fi} channel state information,''
  in \emph{Proc.~ACM WiSec}, 2019, pp. 239--244.

\bibitem{wang2019cross}
W.~Wang, S.~He, L.~Sun, T.~Jiang, and Q.~Zhang, ``Cross-technology
  communications for heterogeneous iot devices through artificial doppler
  shifts,'' \emph{IEEE Trans. Wireless Commun.}, vol.~18, no.~2, pp. 796--806,
  2019.

\bibitem{wang2020enabling}
W.~{Wang}, S.~{He}, Q.~{Zhang}, and T.~{Jiang}, ``Enabling low-power {OFDM} for
  {IoT} by exploiting asymmetric clock rates,'' \emph{IEEE/ACM Trans. Netw.},
  vol.~28, no.~2, pp. 602--611, 2020.

\bibitem{cao2019openpose}
Z.~{Cao}, G.~{Hidalgo Martinez}, T.~{Simon}, S.~{Wei}, and Y.~A. {Sheikh},
  ``{OpenPose}: Realtime multi-person {2D} pose estimation using part affinity
  fields,'' \emph{IEEE Trans. Pattern Anal. Mach. Intell.}, vol.~43, no.~1, pp.
  172--186, 2021.

\bibitem{he2017mask}
K.~He, G.~Gkioxari, P.~Doll{\'a}r, and R.~Girshick, ``{Mask R-CNN},'' in
  \emph{Proc.~IEEE ICCV}, 2017, pp. 2961--2969.

\bibitem{yong2020authenticating}
Y.~Huang, W.~Wang, H.~Wang, T.~Jiang, and Q.~Zhang, ``Authenticating on-body
  iot devices: An adversarial learning approach,'' \emph{IEEE Trans. Wireless
  Commun.}, vol.~19, no.~8, pp. 5234--5245, 2020.

\bibitem{zhao2018rf}
M.~Zhao, Y.~Tian, H.~Zhao, M.~A. Alsheikh, T.~Li, R.~Hristov, Z.~Kabelac,
  D.~Katabi, and A.~Torralba, ``Rf-based {3D} skeletons,'' in \emph{Proc.~ACM
  SIGCOMM}, 2018, p. 267–281.

\bibitem{jiang2020towards}
W.~Jiang, H.~Xue, C.~Miao, S.~Wang, S.~Lin, C.~Tian, S.~Murali, H.~Hu, Z.~Sun,
  and L.~Su, ``Towards {3D} human pose construction using {WiFi},'' in
  \emph{Proc.~ACM MobiCom}, 2020.

\bibitem{wang2019person}
F.~{Wang}, S.~{Zhou}, S.~{Panev}, J.~{Han}, and D.~{Huang}, ``{Person-in-WiFi}:
  Fine-grained person perception using {WiFi},'' in \emph{Proc.~IEEE ICCV},
  2019, pp. 5451--5460.

\bibitem{wang2015resource}
D.~Wang, L.~Toni, P.~C. Cosman, and L.~B. Milstein, ``Resource allocation and
  performance analysis for multiuser video transmission over doubly selective
  channels,'' \emph{IEEE Trans. Wireless Commun.}, vol.~14, no.~4, pp.
  1954--1966, 2015.

\bibitem{adib2015capturing}
F.~Adib, C.-Y. Hsu, H.~Mao, D.~Katabi, and F.~Durand, ``Capturing the human
  figure through a wall,'' \emph{ACM Transactions on Graphics}, vol.~34, no.~6,
  pp. 1--13, 2015.

\bibitem{wang2015understanding}
W.~Wang, A.~X. Liu, M.~Shahzad, K.~Ling, and S.~Lu, ``Understanding and
  modeling of {WiFi} signal based human activity recognition,'' in
  \emph{Proc.~ACM MobiCom}, 2015, pp. 65--76.

\bibitem{ma2019wifi}
Y.~Ma, G.~Zhou, and S.~Wang, ``{WiFi} sensing with channel state information: A
  survey,'' \emph{ACM Comput. Surv.}, vol.~52, no.~3, Jun. 2019.

\bibitem{wang2016gait}
W.~Wang, A.~X. Liu, and M.~Shahzad, ``Gait recognition using wifi signals,'' in
  \emph{Proc.~ACM UbiComp}, 2016, p. 363–373.

\bibitem{ren2017faster}
S.~Ren, K.~He, R.~Girshick, and J.~Sun, ``{Faster R-CNN}: Towards real-time
  object detection with region proposal networks,'' \emph{IEEE Trans. Pattern
  Anal. Mach. Intell.}, vol.~39, no.~6, pp. 1137--1149, 2017.

\bibitem{lin2017feature}
T.-Y. Lin, P.~Doll{\'a}r, R.~Girshick, K.~He, B.~Hariharan, and S.~Belongie,
  ``Feature pyramid networks for object detection,'' in \emph{Proc.~IEEE CVPR},
  2017, pp. 2117--2125.

\bibitem{he2016deep}
K.~He, X.~Zhang, S.~Ren, and J.~Sun, ``Deep residual learning for image
  recognition,'' in \emph{Proc.~IEEE CVPR}, 2016, pp. 770--778.

\bibitem{ji20123d}
S.~Ji, W.~Xu, M.~Yang, and K.~Yu, ``{3D} convolutional neural networks for
  human action recognition,'' \emph{IEEE Trans. Pattern Anal. Mach. Intell.},
  vol.~35, no.~1, pp. 221--231, 2012.

\bibitem{long2015fully}
J.~Long, E.~Shelhamer, and T.~Darrell, ``Fully convolutional networks for
  semantic segmentation,'' in \emph{Proc.~IEEE CVPR}, 2015, pp. 3431--3440.

\bibitem{Halperin2011csitool}
D.~Halperin, W.~Hu, A.~Sheth, and D.~Wetherall, ``Tool release: Gathering
  802.11n traces with channel state information,'' \emph{ACM SIGCOMM CCR},
  vol.~41, no.~1, p.~53, Jan. 2011.

\end{thebibliography}

\end{document}